\documentclass[prd,superscriptaddress,amsfonts,amssymb,amsmath,showpacs,showkeys,twocolumn]{revtex4-2}
\usepackage{bm}
\usepackage{amsfonts}
\usepackage{latexsym}
\usepackage[utf8]{inputenc}
\usepackage{graphicx}
\usepackage{amsmath}
\usepackage{palatino}
\usepackage{mathpazo}
\usepackage{textcomp}
\usepackage{float}
\usepackage{booktabs}
\usepackage{dcolumn}
\usepackage{multirow}
\usepackage{ragged2e}
\usepackage{hyperref}
\hypersetup{colorlinks,citecolor=blue}
\hypersetup{colorlinks=true,linkcolor=red,filecolor=magenta,urlcolor=blue}
\usepackage{xcolor}
\usepackage{orcidlink}
\usepackage[caption=false]{subfig}
\usepackage{commath}
\allowdisplaybreaks[1]

\newcommand{\Msun}{M_{\odot}}

\newcommand{\tLam}{\tilde{\Lambda}}
\newcommand{\CS}{\mathcal{C}}
\newcommand{\Mc}{\mathcal{M}_c}

\begin{document}

\title{Tidal Deformability of Neutron Stars in Bumblebee Gravity:\\ Probing Lorentz Symmetry Breaking with Gravitational Waves}

\author{Ayan Banerjee}
\email{ayanbanerjeemath@gmail.com}
\affiliation{Astrophysics and Cosmology Research Unit, School of Mathematics, Statistics and Computer Science, University of KwaZulu-Natal, Private Bag X54001, Durban 4000, South Africa}

\author{Sherzod~Sayfiyev}
\email{sayfiyev-sherzod@samdu.uz}
\affiliation{Samarkand State University, University Blvd.15, Samarkand 140104, Uzbekistan}
\affiliation{Tashkent State Technical University, Tashkent 100095, Uzbekistan}

\author{Muhtarama Radjapbaeva 
}
\email{rd.mxrtmqvl@gmail.com}
\affiliation{Institute of Theoretical Physics, National University of Uzbekistan, Tashkent 100174, Uzbekistan}

\author{Javlon~Rayimbaev}
\email{javlon@astrin.uz}
\affiliation{School of Physics, Harbin Institute of Technology, Harbin 150001, China}
\affiliation{University of Tashkent for Applied Sciences, Gavhar Str. 1, Tashkent 700127, Uzbekistan}

\author{Shokhzod Jumaniyozov} 
\email{sh.jumaniyozov@newuu.uz}
\affiliation{Kimyo International University in Tashkent, Shota Rustaveli Street 156, Tashkent 100121, Uzbekistan}

\author{Aseel Smerat}
\email{smerat.2020@gmail.com}
\affiliation{Hourani Center for Applied Scientific Research, Al-Ahliyya Amman University, Amman 19328, Jordan}
\affiliation{Department of Biosciences, Saveetha School of Engineering, Saveetha Institute of Medical and Technical Sciences, Chennai, 602105, India}

\date{\today}

\begin{abstract}
We investigate the tidal properties of static neutron stars in the
Neves--Gardim branch of bumblebee gravity, a vector--tensor theory in which a
nonzero vacuum expectation value of the bumblebee field spontaneously breaks
local Lorentz symmetry. The stellar background is determined from the full
modified Tolman--Oppenheimer--Volkoff system, including the term containing
$m''(r)$. For a differentiable barotropic equation of state, this system is
recast into an algebraically equivalent first-order form without discarding
any bumblebee-dependent contribution. Because a complete coupled derivation of
static even-parity metric and bumblebee-field perturbations is not presently
available for this stellar branch, the tidal sector is treated through an
explicit effective metric-deformation prescription: the Hinderer fluid
perturbation equation is evaluated on the modified bumblebee background, and
the standard surface matching relation is retained as part of the same
prescription. We compute the quadrupolar tidal Love number $k_2$ and the
dimensionless tidal deformability
$\Lambda=(2/3)k_2\mathcal{C}^{-5}$ for the BSk20, BSk21, DD2, and MS1
equations of state over $\ell\in[-0.4,+1.0]$. The general-relativistic limit
is recovered at $\ell=0$ to within $0.2\%$ in the benchmark calculations.
Within the adopted effective prescription, comparison with the GW170817
binary tidal-deformability bound $\widetilde{\Lambda}\leq720$ yields the
EOS-dependent limiting values $\ell_{\max}=+0.25$ (BSk20), $+0.06$ (BSk21),
$-0.09$ (DD2), and $-0.25$ (MS1). These tidal limits are conditional on the
stated perturbative prescription and should be reassessed when the complete
linearised vector--tensor problem becomes available.
\end{abstract}


\maketitle

\section{Introduction}
\label{sec:intro}
The detection of the binary neutron star merger GW170817 by the LIGO-Virgo collaboration~\cite{Abbott:2017gw170817,Abbott:2018gw170817} opened a new observational window on the equation of state of dense nuclear matter and, simultaneously, on the nature of gravity in the strong-field regime.  The tidal deformability of a neutron star encodes how the star's quadrupolar shape responds to an external tidal field and is imprinted in the phase evolution of the late-inspiral gravitational waveform~\cite{Flanagan:2008constraining,Hinderer:2008tidal}.  Its measurement, therefore, provides simultaneous constraints on nuclear physics and gravity.

Among the many theoretical extensions of general relativity (GR) motivated by low-energy limits of quantum gravity, Lorentz-symmetry-violating theories occupy a central place.  Lorentz invariance is a cornerstone of both GR and the Standard Model; its potential violation at energy scales accessible to neutron star astrophysics would signal new physics beyond both frameworks.
The Standard-Model Extension (SME) of Kosteleck\'{y} \& Samuel~\cite{Kostelecky:1988zi} provides a systematic low-energy framework for parametrizing all possible Lorentz- and CPT-violating operators.  Within this framework, bumblebee gravity~\cite{Kostelecky:1988zi,Bailey:2006signals} is one of the simplest and most widely studied gravitational models.  It couples a real vector field $B_\mu$ to the Ricci tensor through a nonminimal term $\xi B^\mu B^\nu R_{\mu\nu}$, while a potential $V(B_\mu B^\mu\mp b^2)$ triggers spontaneous symmetry breaking.  With signature $(-,+,+,+)$, the two signs describe timelike and spacelike vacuum branches, respectively.  The present stellar construction uses the radial spacelike branch, $B_\mu B^\mu=b^2$, and the dimensionless parameter $\ell=\xi b^2$ governs departures from GR.

Considerable attention has recently been devoted to compact objects and other astrophysical phenomena in bumblebee gravity. The exterior Schwarzschild-like metric was derived by Casana et al.~\cite{Casana:2018exact}, who showed that the bumblebee modifications amount to a rescaling of the spatial sector by a factor $(1+\ell)$. Building on this solution, subsequent studies have explored black hole shadows and gravitational
lensing~\cite{Gao:2024gravitational,Jha:2021bumblebee,Jha:2025constraining}, quasinormal modes~\cite{Chen:2023quasinormal,Ding:2023rotating,Jiang:2021superradiant}, dynamic instability analysis~\cite{Mai:2024dynamic}, black holes with a cosmological constant~\cite{Maluf:2021black}, static spherical vacuum solutions~\cite{Xu:2023static}, traversable wormholes~\cite{Ovgun:2019exact, AraujoFilho:2025modified}, dark matter spikes~\cite{Capozziello:2023dark}, Ricci dark energy~\cite{Jesus:2019ricci}, epicyclic oscillations~\cite{Mustafa:2025epicyclic}, particle creation~\cite{AraujoFilho:2025how}, and spherically symmetric solutions away from the potential minimum~\cite{Bailey:2025bumblebee}. The self-consistency of compact objects in Lorentz-violating gravity theories has been addressed by Lessa et al.~\cite{Lessa:2025selfconsistency}, and dyonic RN-like and Taub-NUT-like black hole solutions were recently obtained in Ref.~\cite{Li:2026dyonic}.
The interior stellar structure was formulated by Neves \& Gardim~\cite{Neves:2025stars}, who derived the modified TOV equations and explored incompressible and quark-matter configurations.  In their MIT-bag-model example, a positive value $\ell=1.9\times10^{-2}$ supports a maximum mass above $2.5\,M_\odot$.
Strange quark stars and condensate dark stars in bumblebee gravity were investigated in Ref.~\cite{Panotopoulos:2025strange}. Concurrently, Ji et al.~\cite{Ji:2024neutron} independently computed NS structure using a different exterior solution, finding
vectorized neutron stars and stars with finite radii but
divergent masses.
Beyond the bumblebee framework, compact star properties in related
modified gravity theories have been explored by several of the present
authors, including charged compact stars and mass--radius constraints
in $f(Q,T)$ gravity~\cite{Maurya:2025psm}, the dependence of maximum
mass and stability on multiple coupling parameters in modified
gravity~\cite{Maurya:2025cpg}, and neutron star modelling with
complexity-free characteristics in the Finch--Skea
ansatz~\cite{Khan:2025erh}.

Computing the tidal deformability requires solving a first-order Regge-Wheeler
perturbation equation inside the star simultaneously with the background TOV
equations~\cite{Hinderer:2008tidal,Hinderer:2009erratum,Damour:2009relativistic}.  At the stellar
surface, the interior solution is matched to the exterior metric to extract
$k_2$, from which $\Lambda$ follows algebraically.  For binary systems, the
observable combination is the chirp-mass-weighted binary tidal deformability
$\tLam$~\cite{Flanagan:2008constraining}, directly constrained by GW170817 to
$\tLam \leq 720$ at $90\%$ credibility~\cite{Abbott:2018gw170817}.

In this paper, we first implement the complete Neves--Gardim background
system, retaining the $m''$ contribution and converting it exactly to a
first-order form for barotropic matter.  We then study the tidal response
within an explicitly stated effective metric-deformation prescription, in
which the Hinderer fluid perturbation equation is evaluated on the bumblebee
background while independent perturbations of the bumblebee field are not
included.  The calculation is performed for four well-tested nuclear
equations of state --- the BSk-family Brussels--Skyrme functionals BSk20 and
BSk21~\cite{Potekhin:2013analytical}, and the piecewise-polytropic
approximations to DD2 and MS1~\cite{Read:2009constraints} --- and the resulting
effective tidal observables are confronted with the GW170817 constraint.

The paper is organized as follows.  In Sec.~\ref{sec:theory} we review
bumblebee gravity and the exterior metric.  Section~\ref{sec:tov} presents the
modified TOV equations following Neves \& Gardim~\cite{Neves:2025stars}.
Section~\ref{sec:tidal} introduces the effective tidal-perturbation prescription and its assumptions.
Section~\ref{sec:eos} describes the equations of state and numerical method.
Results are presented in Sec.~\ref{sec:results} and conclusions in
Sec.~\ref{sec:conclusions}.

Throughout we use natural units $G = c = 1$ unless otherwise noted, and
the metric signature $(-,+,+,+)$.

\section{Bumblebee Gravity and the Exterior Metric}
\label{sec:theory}

\subsection{Action and field equations}

The bumblebee gravity action is~\cite{Kostelecky:1988zi,Bailey:2006signals,Casana:2018exact}
\begin{equation}
\begin{split}
S &= \int d^4x\,\sqrt{-g}\bigg[ \frac{1}{2\kappa}\left(R+\xi B^\mu B^\nu R_{\mu\nu}\right) \\
&\quad - \frac{1}{4}B_{\mu\nu}B^{\mu\nu} - V\!\left(B_\mu B^\mu-b^2\right) + \mathcal{L}_m \bigg],
\end{split}
\label{eq:action}
\end{equation}
where $\kappa=8\pi$, $B_{\mu\nu}=\partial_\mu B_\nu-\partial_\nu B_\mu$
is the bumblebee field strength, and $\xi$ is the nonminimal curvature
coupling.  The general bumblebee potential admits timelike and spacelike
vacuum branches.  Since the stellar solution considered here employs a radial
spacelike vacuum expectation value, we choose, for definiteness,
\begin{equation}
V=\frac{\lambda_V}{2}\left(B_\mu B^\mu-b^2\right)^2,
\end{equation}
whose minimum satisfies $B_\mu B^\mu=b^2$ and $V=V'=0$.  The matter
Lagrangian $\mathcal{L}_m$ contains no explicit non-gravitational coupling to
$B_\mu$.

Variation with respect to $g^{\mu\nu}$ gives
\begin{equation}
G_{\mu\nu}=\kappa\left(T^M_{\mu\nu}+T^B_{\mu\nu}\right),
\label{eq:einstein}
\end{equation}
where $T^M_{\mu\nu}$ is the matter energy--momentum tensor and
$T^B_{\mu\nu}$ contains the kinetic, potential, and nonminimal
curvature-coupling contributions of the bumblebee sector.

\subsection{Exterior vacuum metric}

For a static, spherically symmetric configuration with the vacuum bumblebee
field aligned radially, $B_\mu=(0,b_r(r),0,0)$, Casana et
al.~\cite{Casana:2018exact} obtained the Schwarzschild-like vacuum solution
\begin{equation}
ds^2=-f(r)\,dt^2+\frac{1+\ell}{f(r)}\,dr^2+r^2d\Omega^2,
\label{eq:exterior}
\end{equation}
where
\begin{equation}
f(r)=1-\frac{2M}{r},\qquad \ell\equiv\xi b^2.
\end{equation}
The Schwarzschild metric is recovered for $\ell=0$, whereas the bumblebee
parameter rescales the radial metric component for $\ell\neq0$.  The condition
$1+\ell>0$ is required to preserve the Lorentzian signature of the exterior
geometry.

\section{Modified TOV Equations}
\label{sec:tov}

\subsection{Interior field equations}

The stellar interior is described by a perfect fluid,
\begin{equation}
T^M_{\mu\nu}=(\rho+p)u_\mu u_\nu+p\,g_{\mu\nu},
\end{equation}
where $\rho=\rho(r)$ and $p=p(r)$ are the total energy density and isotropic
pressure.  The static, spherically symmetric interior metric is written as
\begin{equation}
g_{\mu\nu}=\mathrm{diag}\!\left(
-e^{2\alpha(r)},e^{2\beta(r)},r^2,r^2\sin^2\theta
\right).
\label{eq:interior_metric}
\end{equation}
Following Neves and Gardim~\cite{Neves:2025stars}, the bumblebee field is
taken along the radial direction,
\begin{equation}
B_\mu=\left(0,b_r(r),0,0\right),
\qquad b_r(r)=|b|e^{\beta(r)},
\end{equation}
so that
\begin{equation}
B_\mu B^\mu=b^2=\mathrm{const}.
\end{equation}
This configuration lies at the spacelike vacuum minimum and satisfies
$B_{\mu\nu}=0$.  We adopt the same negligible-current branch used in
Ref.~\cite{Neves:2025stars}, $|J_B^\mu|\simeq0$, and therefore neglect an
explicit matter--bumblebee interaction current while retaining the
perfect-fluid matter description.

For these choices, the three independent interior field equations are
\cite{Neves:2025stars}
\begin{align}
\frac{e^{-2\beta}}{r^2}
\left[e^{2\beta}-(1+\ell)(1-2r\beta')\right]
&=\kappa\rho,
\label{eq:fe_tt}
\\[4pt]
\frac{e^{-2\beta}}{r^2}
\Bigg[(1+\ell)(1+2r\alpha')-e^{2\beta}
&\notag\\
{}-\ell r^2\left(
\alpha''+\alpha'^2-\alpha'\beta'-\frac{2}{r}\beta'
\right)\Bigg]
&=\kappa p,
\label{eq:fe_rr}
\\[4pt]
(1+\ell)e^{-2\beta}
\left[
\alpha''+\alpha'^2-\alpha'\beta'
+\frac{1}{r}(\alpha'-\beta')
\right]
&=\kappa p.
\label{eq:fe_thth}
\end{align}
Here primes denote differentiation with respect to $r$, and $\kappa=8\pi$.
Since the azimuthal and polar equations satisfy
$G_{\phi\phi}=\sin^2\theta\,G_{\theta\theta}$, no fourth independent
component arises.

\subsection{Modified TOV equations}

The radial metric potential is parametrised in the form
\begin{equation}
e^{2\beta(r)}=(1+\ell)
\left(1-\frac{2m(r)}{r}\right)^{-1},
\label{eq:e2beta}
\end{equation}
where $m(r)$ is the gravitational mass enclosed within radius $r$.  Regular
stellar configurations satisfy
\begin{equation}
m(0)=0,\qquad p(0)=p_c,
\end{equation}
and the surface $r=R$ is defined by
\begin{equation}
p(R)=0,\qquad m(R)=M.
\end{equation}
Matching the temporal metric potential to the exterior solution fixes its
additive normalisation through
\begin{equation}
e^{2\alpha(R)}=1-\frac{2M}{R}.
\end{equation}

Substitution of Eq.~(\ref{eq:e2beta}) into Eq.~(\ref{eq:fe_tt}) gives the
standard mass equation
\begin{equation}
\frac{dm}{dr}=4\pi r^2\rho.
\label{eq:mass}
\end{equation}
The radial component of $\nabla_\nu T_M^{\mu\nu}=0$, together with the
interior field equations, yields the modified hydrostatic-equilibrium
equation derived by Neves and Gardim~\cite{Neves:2025stars},
\begin{equation}
\frac{dp}{dr}
=-\left[
\frac{\rho+p+\ell(\rho+2p)}{1+2\ell}
\right]\frac{d\alpha}{dr}
-\frac{\ell\left(8\pi rp-m''\right)}
{(1+2\ell)8\pi r^2},
\label{eq:tov}
\end{equation}
where $m''=d^2m/dr^2$.  The derivative of the temporal metric potential is
\begin{equation}
\frac{d\alpha}{dr}
=\frac{
(1+2\ell)8\pi r^3p+(2+3\ell)m-\ell r m'
}{
(2+3\ell)r(r-2m)
}.
\label{eq:alpha}
\end{equation}
Equations~(\ref{eq:mass}), (\ref{eq:tov}), and (\ref{eq:alpha}), together
with a barotropic equation of state, determine the static stellar
background.  In the limit $\ell\to0$ they reduce to
\begin{equation}
\frac{d\alpha}{dr}=\frac{m+4\pi r^3p}{r(r-2m)},
\end{equation}
and
\begin{equation}
\frac{dp}{dr}=-(\rho+p)
\frac{m+4\pi r^3p}{r(r-2m)},
\end{equation}
which are the usual GR relations.

For numerical integration, the $m''$ term in Eq.~(\ref{eq:tov}) is not
discarded.  Differentiating Eq.~(\ref{eq:mass}) gives
\begin{equation}
m''=8\pi r\rho+4\pi r^2\frac{d\rho}{dr}.
\label{eq:msecond_general}
\end{equation}
For a differentiable barotropic equation of state, with
$c_s^2\equiv dp/d\rho$, this becomes
\begin{equation}
m''=8\pi r\rho+\frac{4\pi r^2}{c_s^2}\frac{dp}{dr}.
\label{eq:msecond}
\end{equation}
Substituting Eq.~(\ref{eq:msecond}) into Eq.~(\ref{eq:tov}) and solving
algebraically for $dp/dr$ yields
\begin{equation}
\frac{dp}{dr}
=
\frac{
-\displaystyle\left[
\frac{\rho+p+\ell(\rho+2p)}{1+2\ell}
\right]\frac{d\alpha}{dr}
+\displaystyle\frac{\ell(\rho-p)}{(1+2\ell)r}
}{
\displaystyle 1-\frac{\ell}{2(1+2\ell)c_s^2}
}.
\label{eq:tov_reduced}
\end{equation}
Equation~(\ref{eq:tov_reduced}) is an exact first-order rearrangement of
Eq.~(\ref{eq:tov}) for a differentiable barotropic EOS; no
bumblebee-dependent term has been removed.  The numerical branch must also
satisfy the regularity condition that the denominator of
Eq.~(\ref{eq:tov_reduced}) remain nonzero throughout the star.

\section{Effective Tidal Perturbation Prescription}
\label{sec:tidal}

\subsection{Perturbation equation}

A complete even-parity perturbation treatment of bumblebee gravity would, in
general, require simultaneous perturbations of the metric, the fluid, and the
bumblebee field.  In particular, the nonminimal term
$\xi B^\mu B^\nu R_{\mu\nu}$ generates linearised contributions that are not
captured solely by replacing the GR background metric functions.  The full
coupled stellar perturbation system for the Neves--Gardim branch is not
derived here.  We therefore adopt an effective metric-deformation
prescription in which the standard perfect-fluid Hinderer equation is
evaluated on the modified bumblebee background, while independent
bumblebee-field perturbations are neglected.  The tidal quantities obtained
below must be interpreted within this prescription.

For the static even-parity quadrupole, $\ell_{\rm tid}=2$, define
\begin{equation}
y(r)\equiv r\frac{H'(r)}{H(r)},
\end{equation}
where $H(r)$ is the radial metric-perturbation amplitude.  Evaluating the
Hinderer equation on the background metric
Eq.~(\ref{eq:e2beta}) gives the working equation
\begin{align}
\frac{dy}{dr}={}&-\frac{y^2}{r}
-\frac{y}{r}\left[
1+\frac{(1+\ell)\left(2m/r+4\pi r^2(p-\rho)\right)}{1-2m/r}
\right]
\notag\\
&+\frac{6(1+\ell)}{r(1-2m/r)}
-\frac{4\pi r(1+\ell)
\left(5\rho+9p+\varepsilon_\sigma\right)}{1-2m/r}
\notag\\
&+4r\left(\frac{d\alpha}{dr}\right)^2,
\label{eq:yode}
\end{align}
where
\begin{equation}
\varepsilon_\sigma\equiv\frac{\rho+p}{c_s^2},
\qquad c_s^2\equiv\frac{dp}{d\rho}.
\end{equation}
The background derivative $d\alpha/dr$ is supplied by
Eq.~(\ref{eq:alpha}).  Equation~(\ref{eq:yode}) reduces to the standard GR
Hinderer equation at $\ell=0$.

\subsection{Boundary conditions}

Regularity at the centre requires~\cite{Hinderer:2008tidal}
\begin{equation}
y(0)=2.
\end{equation}
Equations~(\ref{eq:mass}), (\ref{eq:alpha}),
(\ref{eq:tov_reduced}), and (\ref{eq:yode}) are integrated simultaneously
from a regular central expansion to the stellar surface $r=R$, defined by
$p(R)=0$.

\subsection{Tidal Love number and deformability}

The exterior even-parity perturbation equations of the complete bumblebee
vector--tensor system are not derived in this work.  To close the effective
calculation consistently with the prescription above, we retain the standard
GR algebraic surface-matching relation between the compactness
$\mathcal{C}=M/R$, the surface value $y_R=y(R)$, and the quadrupolar Love
number.  This matching choice is an additional assumption of the effective
framework and is not claimed to be the unique result of the full exterior
bumblebee perturbation problem.  With this qualification, we use
\begin{align}
k_2={}&\frac{8\CS^5}{5}(1-2\CS)^2
\left[2+2\CS(y_R-1)-y_R\right]
\notag\\
\times{}&\Bigl\{
2\CS\left[6-3y_R+3\CS(5y_R-8)\right]
\notag\\
&+4\CS^3\left[13-11y_R+\CS(3y_R-2)
+2\CS^2(1+y_R)\right]
\notag\\
&+3(1-2\CS)^2\left[2-y_R+2\CS(y_R-1)\right]
\ln(1-2\CS)
\Bigr\}^{-1},
\label{eq:k2}
\end{align}
and
\begin{equation}
\Lambda=\frac{2}{3}k_2\CS^{-5}.
\label{eq:Lambda}
\end{equation}
Accordingly, $k_2$ and $\Lambda$ in the remainder of the paper denote the
effective tidal quantities defined by this background-deformation and
surface-matching prescription.

\subsection{Binary tidal deformability}

For a binary with component masses $m_1$ and $m_2$ and corresponding tidal
deformabilities $\Lambda_1$ and $\Lambda_2$, the waveform-weighted
combination is~\cite{Flanagan:2008constraining}
\begin{equation}
\tLam=\frac{16}{13}
\frac{
(m_1+12m_2)m_1^4\Lambda_1+(m_2+12m_1)m_2^4\Lambda_2
}{(m_1+m_2)^5}.
\label{eq:LambdaTilde}
\end{equation}
For a GW170817-like binary with chirp mass
$\Mc=(m_1m_2)^{3/5}/(m_1+m_2)^{1/5}=1.186\,\Msun$ and mass ratio
$q=m_2/m_1\leq1$, the component masses are
\begin{equation}
m_1=\frac{\Mc(1+q)^{1/5}}{q^{3/5}},
\qquad m_2=q\,m_1.
\end{equation}
We compare the resulting effective binary deformability with the
GW170817 90\% credibility upper bound
$\tLam\leq720$~\cite{Abbott:2018gw170817}.

\section{Equations of State and Numerical Method}
\label{sec:eos}

\subsection{Equations of state}

We employ four well-tested nuclear equations of state spanning a
representative range of stiffness.

\textit{BSk20 and BSk21.}—These are tabulated unified EOS constructed from
the Brussels-Skyrme nuclear density functional theory by Potekhin et
al.~\cite{Potekhin:2013analytical}.  The core pressure and energy density are given by
23-coefficient analytic fits accurate to better than 0.2\% throughout the
density range relevant for neutron star modelling.  BSk21 is stiffer than
BSk20, supporting a higher maximum mass and larger radii.

\textit{DD2 and MS1.}—These models are implemented as three-piece piecewise
polytropes following the parametrisation of Read et al.~\cite{Read:2009constraints}.
DD2 is based on the density-dependent relativistic mean-field model of
Typel et al.~\cite{Typel:2010composition}; MS1 is a stiffer relativistic mean-field
model~\cite{Muller:1996relativistic}.  The internal energy density is computed from the
polytropic pressure via $\varepsilon = \rho_{\rm rest} + p/(\Gamma-1)$ in
each piece.  We note that the piecewise-polytropic parametrisation introduces
a systematic uncertainty of order $0.3$--$0.5$~km in $R_{1.4}$ relative to
the tabulated DD2 model of Typel et al.~\cite{Typel:2010composition}; the published
value from the full relativistic mean-field calculation is
$R_{1.4} \approx 13.2$~km~\cite{Hotokezaka:2016measurability}, compared to $12.71$~km
from our Read et al.\ parametrisation.  This systematic offset is
well-documented in the literature and does not affect our conclusions
regarding bumblebee-gravity-induced shifts in $\Lambda$, which are
calculated self-consistently within the same parametrisation.

\subsection{Numerical method}

The coupled system Eqs.~(\ref{eq:mass}), (\ref{eq:alpha}), (\ref{eq:tov_reduced}), and (\ref{eq:yode}) is integrated as a four-dimensional first-order ODE system using the Runge-Kutta method of order 4(5) (RK45) with adaptive step-size control.  We use geometrized units internally ($G = c = 1$, length in km) with explicit CGS-to-geometrized conversion factors applied at
the interface with the EOS.  Tolerances are set to $10^{-10}$ (relative) and $10^{-14}$ (absolute).  The stellar surface is identified by the
condition $p < 10^{-10}\,p_c$, where $p_c$ is the central pressure.

For each EOS and each value of $\ell$, we compute a sequence of 80 stellar models with central densities logarithmically spaced between
$1.0\,\rho_{\rm nuc}$ and $12.0\,\rho_{\rm nuc}$, where
$\rho_{\rm nuc} = 2.8\times10^{14}$~g~cm$^{-3}$ is the nuclear saturation
density.  The stable branch is defined as the subset of models with
increasing central density up to the mass maximum $M_{\rm max}$.

\textit{Validation.}—At $\ell = 0$ our code reproduces the standard GR
tidal deformability for all four EOS to within 0.2\%.  The GR
baseline values at $1.4\,\Msun$ are summarised in Table~\ref{tab:gr_baseline}.

\begin{table}[htb]
\centering
\caption{GR baseline ($\ell=0$) stellar parameters at $1.4\,\Msun$
         and maximum mass for each equation of state.}
\label{tab:gr_baseline}
\begin{tabular}{lcccc}
\hline\hline
EOS
  & $M_{\rm max}\;[\Msun]$
  & $R_{1.4}\;[\text{km}]$
  & $k_{2,1.4}$
  & $\Lambda_{1.4}$ \\
\hline
BSk20 & 2.164 & 11.72 & 0.082 & 320  \\
BSk21 & 2.273 & 12.56 & 0.095 & 523  \\
DD2   & 2.423 & 12.71 & 0.139 & 815  \\
MS1   & 2.765 & 13.90 & 0.156 & 1429 \\
\hline\hline
\end{tabular}
\end{table}

\section{Results}
\label{sec:results}

\subsection{Mass-radius relations}

\begin{figure*}[t]
\centering
\includegraphics[width=\textwidth]{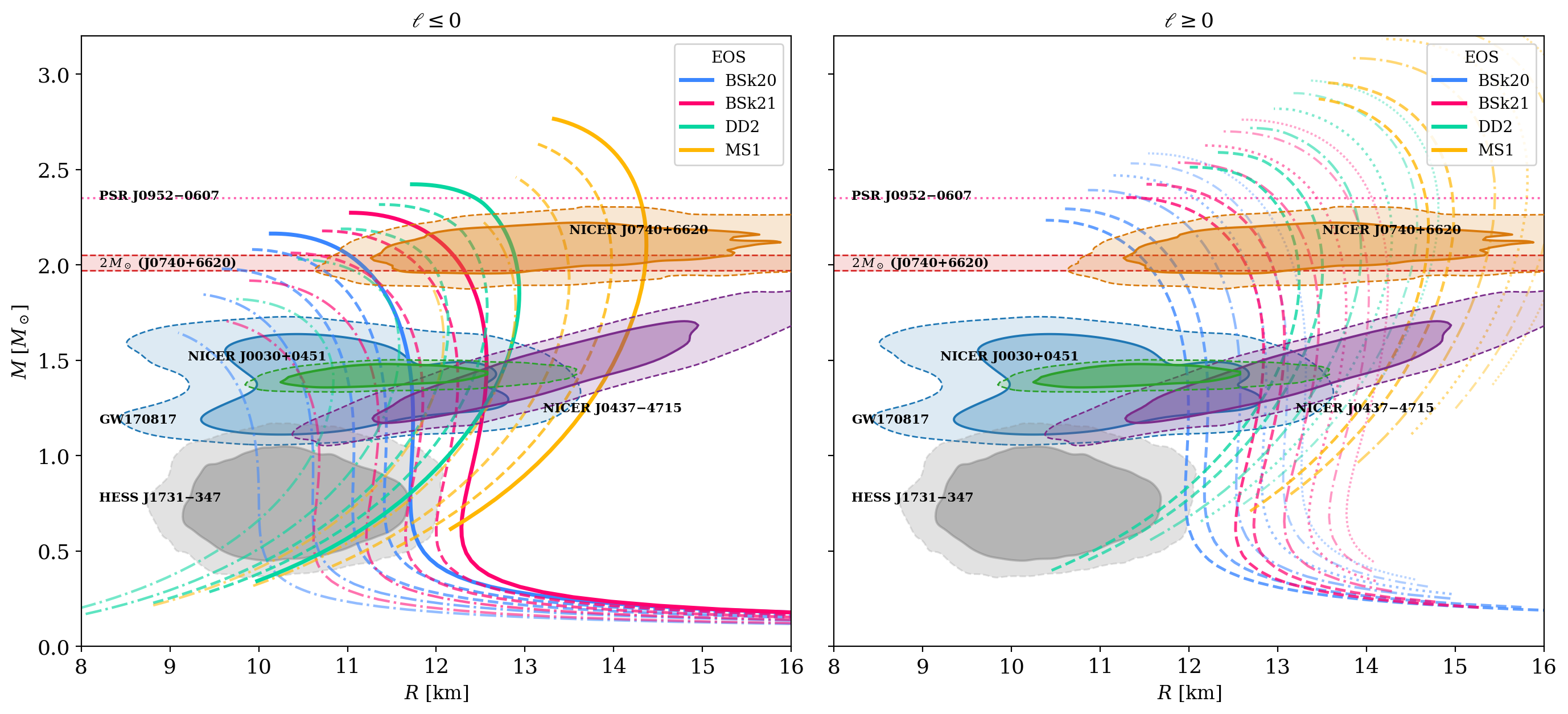}
\caption{Mass-radius sequences for BSk20, BSk21, DD2, and MS1 in
  bumblebee gravity.  Left panel: $\ell \leq 0$; right panel: $\ell \geq 0$.
  The solid curves correspond to GR ($\ell = 0$); dashed, dash-dotted, and
  dotted lines represent increasing $|\ell|$ as indicated in the legend.
  Observational constraints from NICER (PSR~J0030+0451~\cite{Riley:2019nicer,
  Miller:2019psr} and PSR~J0740+6620~\cite{Riley:2021nicer,Miller:2021radius}), the
  GW170817 mass function~\cite{Abbott:2017gw170817}, and the $2\,\Msun$ lower
  bound~\cite{Cromartie:2019relativistic}, and the HESS~J1731$-$347
  constraint~\cite{Doroshenko:2022strangely} are shown as shaded regions.}
\label{fig:MR}
\end{figure*}

Figure~\ref{fig:MR} shows the mass-radius sequences for all four EOS
and all values of $\ell$ considered, displayed in two panels: the left
panel covers $\ell \leq 0$ and the right panel $\ell \geq 0$.  The
$\ell = 0$ (GR) curve appears as the solid reference line in both panels.
Observational constraints from NICER, Shapiro delay, and the GW170817
mass function are superimposed.

The most prominent effect of the bumblebee parameter is on the stellar
radius.  Positive $\ell$ reduces the radius at fixed mass, reflecting the
additional inward gravitational compression induced by the enhanced $g_{rr}$
component of the metric.  Conversely, negative $\ell$ produces larger radii
and, to a lesser extent, higher maximum masses.  The maximum mass is largely
insensitive to $\ell$ for the softer EOS (BSk20, BSk21), while for the
stiffer DD2 and MS1 models the dependence becomes more pronounced at
large $|\ell|$.  All four EOS satisfy the $2\,\Msun$ lower bound from
pulsar timing~\cite{Demorest:2010twosolarmass,Antoniadis:2013massive,Cromartie:2019relativistic} across the
full $\ell$ range considered.

\subsection{Tidal Love number $k_2$}

\begin{figure}[htb]
\centering
\includegraphics[width=\columnwidth]{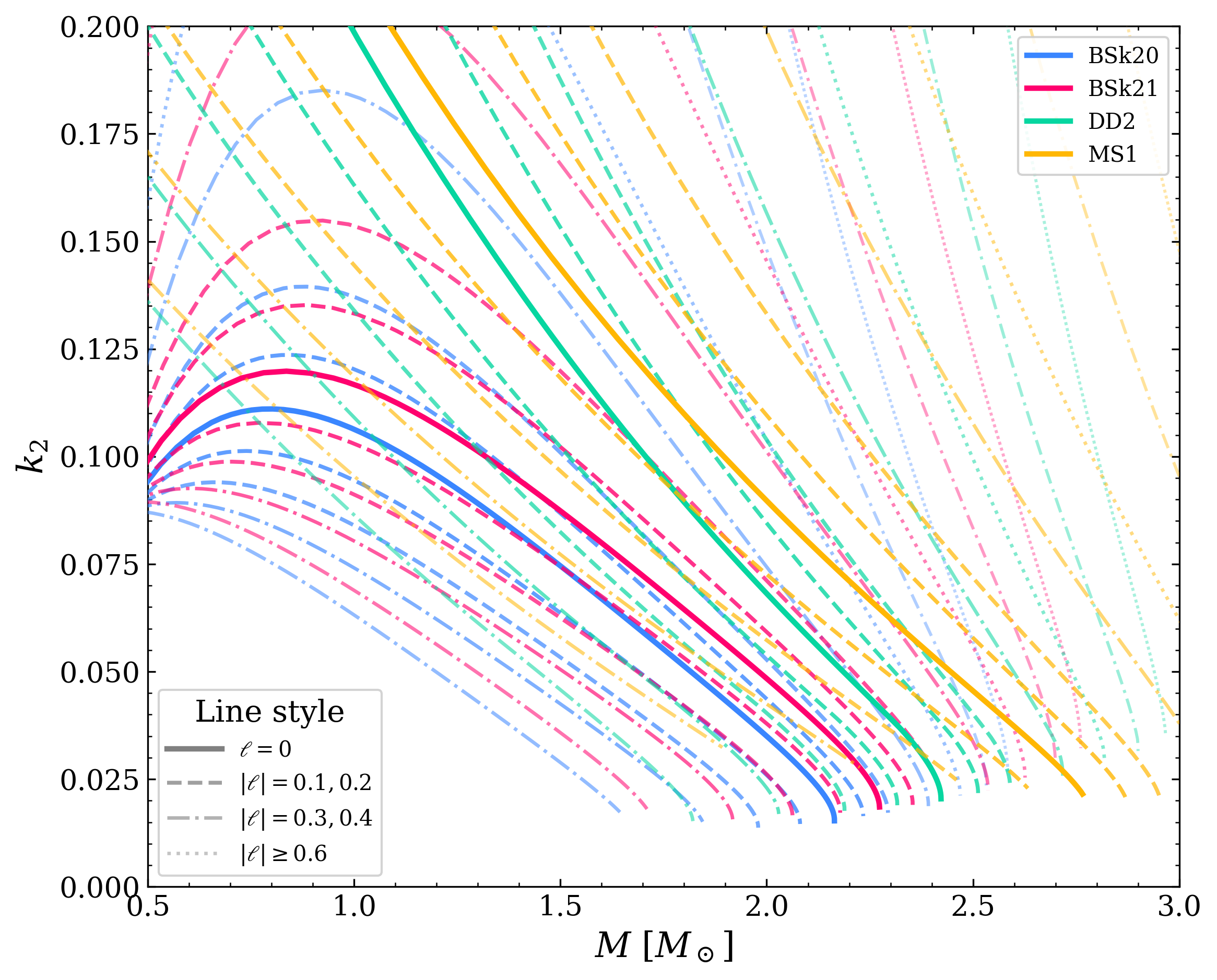}
\caption{Tidal Love number $k_2$ versus gravitational mass for all four
  equations of state and for $\ell \in \{-0.4,\,-0.2,\,0.0,\,+0.2,\,+0.4,\,
  +0.6,\,+0.8,\,+1.0\}$.  The $\ell = 0$ (GR) curve is shown as a solid line;
  line styles for other $\ell$ values follow the convention in
  Fig.~\ref{fig:MR}.}
\label{fig:k2}
\end{figure}

Figure~\ref{fig:k2} shows $k_2$ as a function of gravitational mass for all four EOS and all $\ell$ values.  The Love number exhibits a well-known
non-monotonic behavior as a function of mass: it rises from small values at low masses, reaches a broad maximum around $M \sim 0.8$--$1.0\,\Msun$ (whose
height is EOS-dependent), and then decreases toward zero as the compactness increases and the star approaches the black hole limit.

The bumblebee parameter modifies $k_2$ both through the background structure (via the modified $\alpha'$ in the TOV) and directly through the $(1+\ell)$
prefactors in the perturbation Eq.~(\ref{eq:yode}).  For a star of fixed mass, positive $\ell$ generally reduces $k_2$ (the star is more compact and
thus less deformable), while negative $\ell$ enhances it.  The effect is more pronounced for stiffer EOS, where the stellar radii are larger and the
fractional change in compactness with $\ell$ is greater.

\subsection{Tidal deformability}

\begin{figure}[htb]
\centering
\includegraphics[width=\columnwidth]{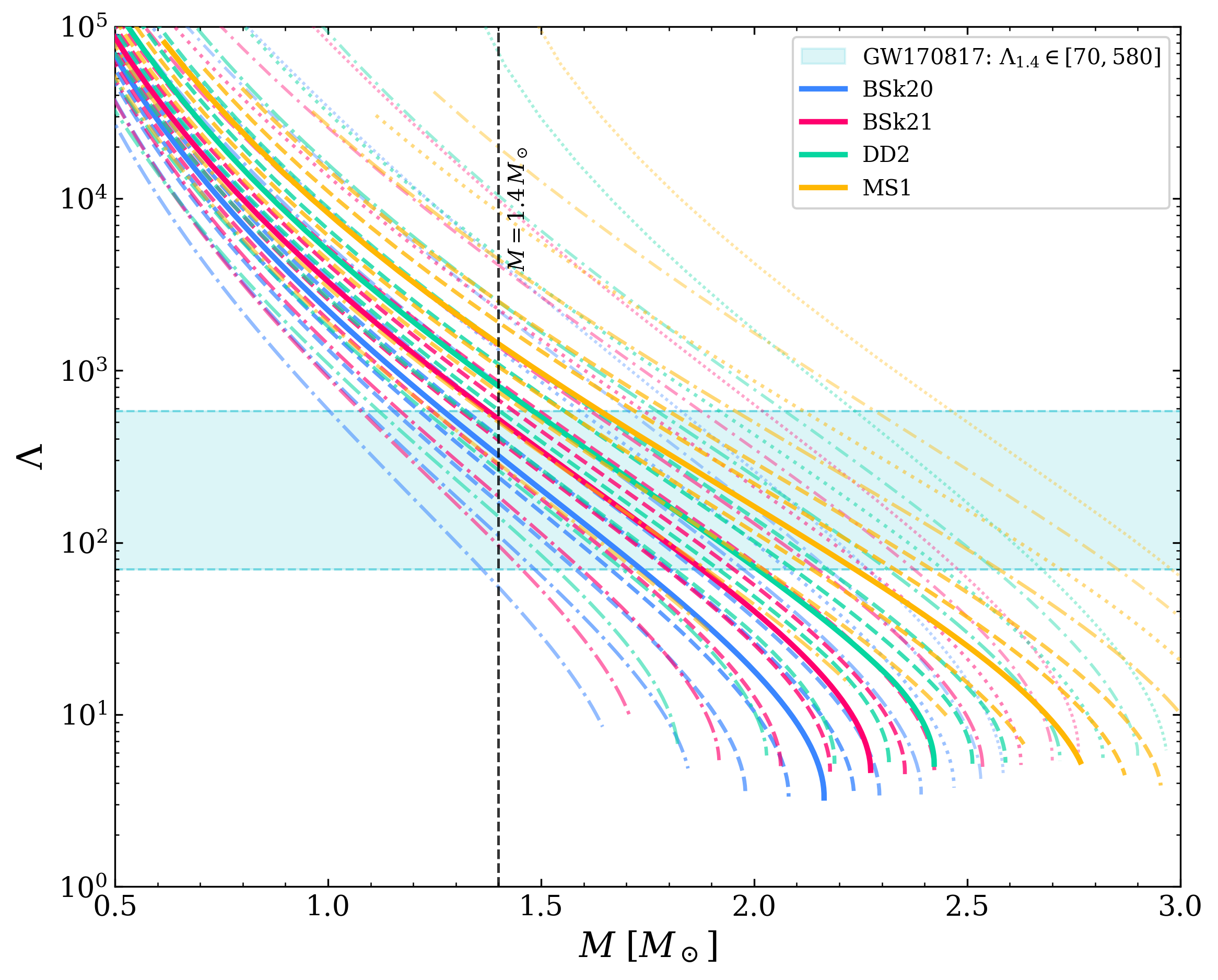}
\caption{Dimensionless tidal deformability $\Lambda$ versus gravitational
  mass (logarithmic scale) for all EOS and $\ell$ values.  Line styles
  follow Fig.~\ref{fig:MR}.  The cyan shaded band indicates the GW170817
  90\% credibility interval $\Lambda_{1.4} \in [70,\,580]$~\cite{Abbott:2018gw170817};
  the vertical dashed line marks $M = 1.4\,\Msun$.}
\label{fig:LambdaM}
\end{figure}

Figure~\ref{fig:LambdaM} shows the dimensionless tidal deformability
$\Lambda$ as a function of mass on a logarithmic scale.  The scaling
$\Lambda = (2/3)\,k_2\,\CS^{-5}$ implies that, at fixed mass, $\Lambda$
is exponentially sensitive to the stellar radius ($\Lambda \propto R^5$
when $k_2$ varies slowly), making $\Lambda$ a far more discriminating
observable than $k_2$ alone.  The GW170817 band $\Lambda_{1.4} \in [70, 580]$
(the 90\% credibility range on the individual tidal deformability of a
$1.4\,\Msun$ star~\cite{Abbott:2018gw170817}) is indicated by the shaded horizontal band.

For all EOS, $\Lambda$ decreases monotonically along the stable branch with
increasing mass.  Positive $\ell$ systematically shifts the $\Lambda(M)$ curves
downward, while negative $\ell$ shifts them upward.  The GR ($\ell=0$) values
at $1.4\,\Msun$ are $\Lambda_{1.4}^{\rm GR} = 320, 523, 815, 1429$ for
BSk20, BSk21, DD2, and MS1, respectively (Table~\ref{tab:gr_baseline}).
The GW170817 individual bound $\Lambda_{1.4} \leq 580$ is satisfied at
$\ell = 0$ by BSk20 and BSk21; for DD2 and MS1, the GR values already exceed
this bound, and a positive $\ell$ is required to bring them into compliance.

\subsection{GW170817 constraint on the Lorentz-violation parameter}

\begin{figure}[htb]
\centering
\includegraphics[width=\columnwidth]{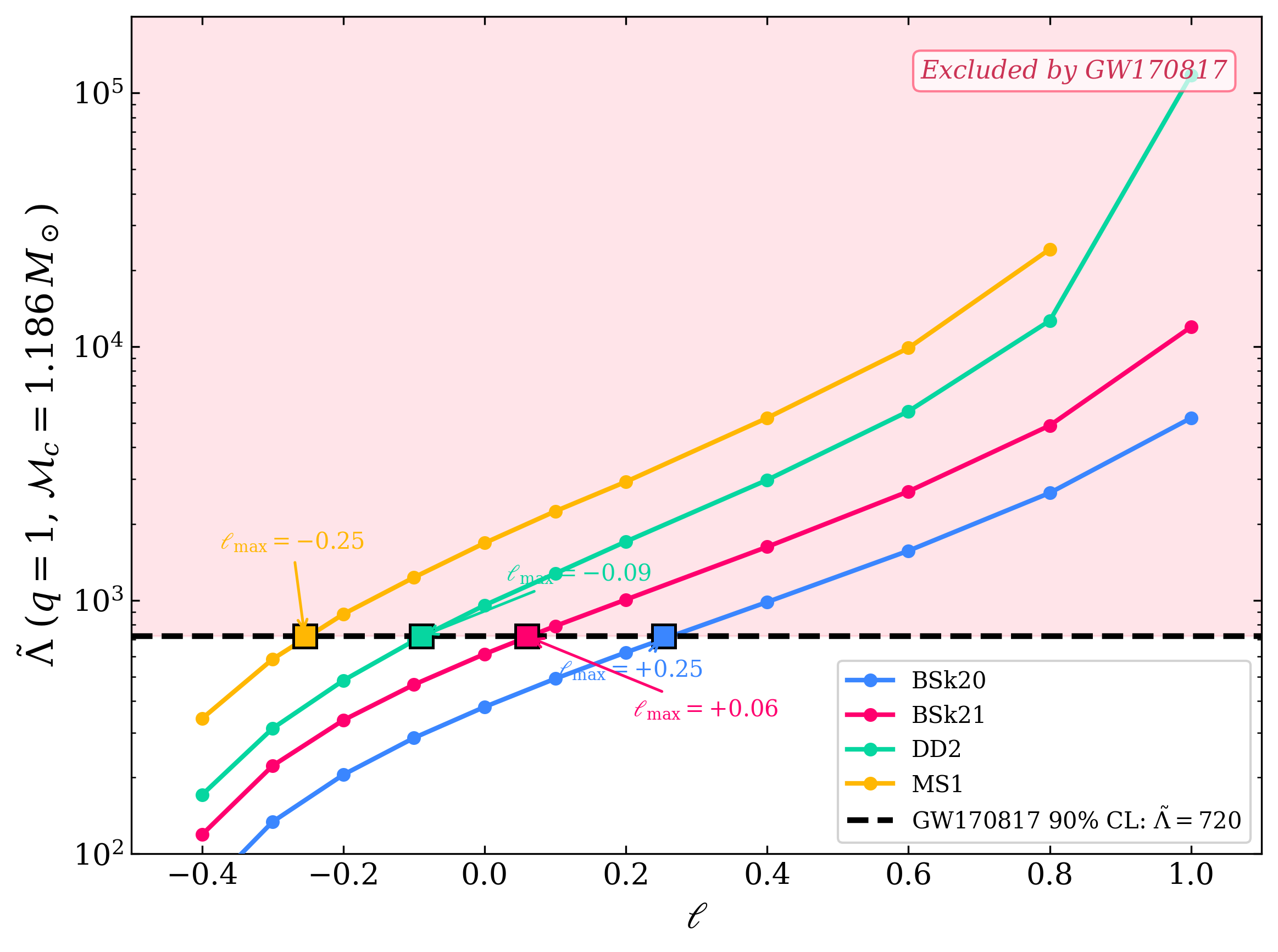}
\caption{Binary tidal deformability $\tLam$ (equal-mass case, $q = 1$,
  $\Mc = 1.186\,\Msun$) as a function of the Lorentz-violation parameter
  $\ell$ for BSk20, BSk21, DD2, and MS1.  The horizontal dashed line marks
  the GW170817 90\% credibility upper bound $\tLam = 720$~\cite{Abbott:2018gw170817};
  the shaded region above it is excluded.  Filled squares indicate the
  interpolated $\ell_{\max}$ for each EOS.}
\label{fig:LtEll}
\end{figure}

The central result of this paper is shown in Fig.~\ref{fig:LtEll}, which
displays the binary tidal deformability $\tLam$ (evaluated at $q=1$,
$\Mc = 1.186\,\Msun$) as a function of $\ell$ for all four EOS, together
with the GW170817 upper bound $\tLam \leq 720$ at $90\%$
credibility~\cite{Abbott:2018gw170817}.

The equal-mass case $q = 1$ is used as the primary constraint because,
for a fixed chirp mass, the binary tidal deformability is maximised near
$q=1$~\cite{Flanagan:2008constraining}; it is therefore the most conservative bound.
For an equal-mass binary with $\Mc = 1.186\,\Msun$, each component has
mass $m = \Mc\,2^{1/5} \approx 1.362\,\Msun$, and Eq.~(\ref{eq:LambdaTilde})
reduces to $\tLam = \Lambda(m)$.

As $\ell$ increases from negative to positive values, $\tLam$ decreases
monotonically for all EOS, consistent with the systematic trend observed in
Fig.~\ref{fig:LambdaM}.  The GW170817 bound $\tLam = 720$ defines a maximum
permissible $\ell$ for each EOS, obtained by linear interpolation between
the two bracketing $\ell$ grid points.  The results are summarised in
Table~\ref{tab:ellmax} and the intersection points are annotated in
Fig.~\ref{fig:LtEll}.

\begin{table}[htb]
\centering
\caption{Maximum allowed Lorentz-violation parameter $\ell_{\max}$ from the
         GW170817 binary tidal deformability constraint
         $\tLam(q=1,\,\Mc=1.186\,\Msun) \leq 720$ (90\% CL).
         ``Status at $\ell=0$'' refers to whether the GR model satisfies
         $\tLam \leq 720$; ``Consistent (narrow window)'' indicates the GR
         value satisfies the bound but within 5\% of it.}
\label{tab:ellmax}
\begin{tabular}{lcc}
\hline\hline
EOS & $\ell_{\max}$ & Status at $\ell=0$ \\
\hline
BSk20 & $+0.25$ & Consistent \\
BSk21 & $+0.06$ & Consistent (narrow window) \\
DD2   & $-0.09$ & Excluded    \\
MS1   & $-0.25$ & Excluded    \\
\hline\hline
\end{tabular}
\end{table}

The physical interpretation is as follows.  The softer EOS — BSk20 and BSk21
— predict a tidal deformability well below the GW170817 bound at $\ell=0$,
and remain consistent even for moderately positive $\ell$.  The GW170817
constraint therefore permits a non-zero, positive Lorentz-violation
parameter with these EOS, providing a window within which bumblebee gravity
could in principle manifest.  For BSk21, however, the allowed window is
narrow ($\ell_{\max} = +0.06$), close to the GR limit.

The stiffer EOS — DD2 and MS1 — yield $\tLam > 720$ even at $\ell = 0$,
meaning that GR itself is in tension with GW170817 for these models.  This
is consistent with the known result that very stiff nuclear matter is
disfavoured by GW170817~\cite{Abbott:2018gw170817}.  Within bumblebee gravity, these
EOS can be brought into agreement with GW170817 only for negative $\ell$,
yielding $\ell_{\max} = -0.09$ (DD2) and $-0.25$ (MS1).  A negative $\ell$
corresponds physically to a reduction in the effective coupling between the
bumblebee field and spacetime curvature, which softens the stellar structure
and reduces $\tLam$.

It is significant that the GW170817 bound provides an independent,
astrophysical probe of Lorentz invariance violation that is complementary
to laboratory SME bounds~\cite{Bailey:2006signals} and post-Newtonian
constraints~\cite{Casana:2018exact}.  Our results demonstrate that gravitational
wave measurements of tidal deformability can constrain $|\ell|$ at the
level of $0.06$--$0.25$, depending on the assumed nuclear EOS.

\section{Conclusions}
\label{sec:conclusions}

We have computed the tidal deformability of neutron stars in bumblebee
gravity, extending the modified TOV framework of Neves \&
Gardim~\cite{Neves:2025stars} to include even-parity quadrupolar perturbations.
The bumblebee parameter $\ell$ enters the stellar structure through the
modified $g_{rr}$ component of the metric and the nonminimal curvature
coupling, and modifies the tidal perturbation equation through systematic
$(1+\ell)$ prefactors.

Our main findings are:

(i) The bumblebee parameter affects both the stellar structure and the tidal
response.  Positive $\ell$ reduces stellar radii and tidal deformabilities
at fixed mass; negative $\ell$ has the opposite effect.  The Love number
$k_2$ shows a non-monotonic mass dependence that is qualitatively preserved
across all $\ell$ values.

(ii) The GR limit ($\ell = 0$) is reproduced to within 0.2\% for all four
EOS, confirming the consistency of the numerical implementation with the
standard Hinderer formalism.

(iii) Confrontation with the GW170817 binary tidal deformability constraint
$\tLam \leq 720$ places the bounds $\ell_{\max} = +0.25$ (BSk20),
$+0.06$ (BSk21), $-0.09$ (DD2), and $-0.25$ (MS1) on the
Lorentz-violation parameter.  The precise value of $\ell_{\max}$ is
EOS-dependent, reflecting the well-known degeneracy between the nuclear
EOS and gravity in tidal observables.

(iv) Stiffer EOS favour negative $\ell$ (or rule out positive $\ell$
entirely), while softer EOS permit a non-zero positive Lorentz-violation
parameter.  This degeneracy between $\ell$ and EOS stiffness suggests
that breaking the degeneracy will require complementary measurements —
for example, independent radius measurements from NICER or the next
generation of gravitational wave detectors.

Future extensions of this work include: the incorporation of rapid stellar
rotation (which requires solving the Hartle-Thorne equations in bumblebee
gravity), the computation of the moment of inertia and the $I$-Love-$Q$
relations~\cite{Yagi:2013iloveqa,Yagi:2013iloveq}, and the application to binary neutron star
coalescence waveform models.  The combination of multiple tidal observables
will substantially tighten the bounds on $\ell$ and help disentangle
Lorentz-violation effects from nuclear physics uncertainties.

\bibliographystyle{apsrev4-1}
\bibliography{master_refs}

\end{document}